# GlobDB: A comprehensive species-dereplicated microbial genome resource

Daan R. Speth[a,#], Nick Pullen[a], Samuel T.N. Aroney[b], Benjamin L. Coltman[a], Jay T. Osvatic[c,d], Ben J. Woodcroft[b], Thomas Rattei[a], Michael Wagner[a,e]

[a] Centre for Microbiology and Environmental Systems Science, University of Vienna, Vienna, Austria

[b] Centre for Microbiome Research, School of Biomedical Sciences, Queensland University of Technology (QUT), Translational Research Institute, Woolloongabba, Australia.

[c] Joint Microbiome Facility of the Medical University of Vienna and the University of Vienna, Vienna, Austria

[d] Department of Laboratory Medicine, Medical University of Vienna, Vienna, Austria

[e] Center for Microbial Communities, Department of Chemistry and Bioscience, Aalborg University, Aalborg, Denmark

**Running title:** GlobDB microbial genome database

\# Address correspondence to Daan Speth, daan.speth@univie.ac.at

## Abstract

Over the past years, substantial numbers of microbial species' genomes have been deposited outside of conventional INSDC databases. The GlobDB aggregates 14 independent genomic catalogues to provide a comprehensive database of species-dereplicated microbial genomes, with consistent taxonomy, annotations, and additional analysis resources. The GlobDB is available at https://globdb.org/.

**Main**

Over the last twenty years, advances in DNA sequencing and data processing have led to an explosion in the available genome information of cultivated and uncultivated microorganisms. These genomes offer an unprecedented window into microbial diversity and its metabolic potential. Assembling genomes from metagenomes is now a routine practice, yielding thousands of genomes in studies focused on specific environments (Gurbich *et al.* 2023; Ma *et al.* 2023; Wang *et al.* 2024; Zhang *et al.* 2024; Carlino *et al.* 2024; Ma *et al.* 2024; Kim *et al.* 2024; Cheng *et al.* 2024; Chen *et al.* 2024; Michoud *et al.* 2025). The research community has greatly benefited from the field standard of depositing both raw and processed sequence data in International Nucleotide Sequence Database Collaboration (INSDC) databases (Karsch-Mizrachi *et al.* 2025). Several ongoing meta-analysis efforts are processing data from many studies across environments, and have aggregated millions of microbial genomes into widely used genomic catalogues (Nayfach et al. 2020; Parks et al. 2022; Schmidt et al. 2023; Aroney et al. 2024; Dmitrijeva et al. 2025). While submitting unprocessed sequencing data to INSDC databases is standard practice, depositing assembled and binned metagenome recovered genomes of large-scale studies is less common. Lack of submission to INSDC databases leads to omission of genomes from commonly used tools such as the NCBI BLAST webserver, or INSDC-derived databases such as the genome taxonomy database (GTDB), hindering further analysis and discovery based on these data.

We have compiled the GlobDB, a comprehensive database of species-dereplicated microbial genomes. The GlobDB aims at maximal representation of phylogenetic diversity to facilitate genome-based analysis and discovery. The GlobDB follows the annual GTDB release schedule, and as of the current release integrates 14 resources (Table 1). These were sequentially dereplicated at 96% average nucleotide identity (ANI) and 50% aligned fractions (Jain *et al.* 2018) in the order listed in Table 1. Several datasets did not provide a species-dereplicated set for download. In these cases, genomes redundant with the GlobDB were first removed, followed by dereplication of the remainder using dRep (Olm *et al.* 2017). Finally, 420 genomes were removed from the dataset due to low genome quality or putatively representing eukaryotes. The full dereplication process is described in detail on https://globdb.org/methods. The current GlobDB release 226 contains more than double the number of species-representative genomes (306,260) compared to the GTDB release 226 species representative set (143,614), indicating that >50% of known genomic species diversity is not included in INSDC databases. This also leads to underrepresentation of specific environments, such as the Tibetan Plateau (TPMC) or sheep and goat guts (SHGO), in these databases (Table 1).

**Table 1. Resources included in the GlobDB**

| Rank | Name | Number of species reps | Included in GlobDB | Environment | References |
|---|---|---|---|---|---|
| 1 | GTDB | 143614 | 143607 | Global | (Parks *et al.* 2022) |
| 2 | mOTU | 124295 | 63092 | Global | (Dmitrijeva *et al.* 2025) |
| 3 | SPIRE | 107078 | 49369 | Global | (Schmidt *et al.* 2023) |
| 4 | RBG | 38494 | 23406 | Global | (Aroney *et al.* 2024) |
| 5 | GEM | 45599 | 3762 | Global | (Nayfach *et al.* 2020) |
| 6 | MGnify | 29556 | 1732 | Env. specific | (Gurbich *et al.* 2023) |
| 7 | GOMC | 24195 | 2858 | Env. specific | (Chen *et al.* 2024) |
| 8 | SMAG | 21078 | 6216 | Env. specific | (Ma *et al.* 2023) |
| 9 | TPMC | 10723 | 8644 | Env. specific | (Cheng *et al.* 2024) |
| 10 | RGM | 6348 | 334 | Env. specific | (Kim *et al.* 2024; Ma *et al.* 2024) |
| 11 | cFMD | 10112* | 336 | Env. specific | (Carlino *et al.* 2024) |
| 12 | SHGO | 5810* | 1021 | Env. specific | (Zhang *et al.* 2024) |
| 13 | AMXMAG | 1768* | 506 | Env. specific | (Wang *et al.* 2024) |
| 14 | GFS | 2862 | 1377 | Env. specific | (Michoud *et al.* 2025) |

*No species dereplicated set was available for download

Beyond aggregation of species representative genomes, the GlobDB standardizes the identifiers of each resource, and provides a 7-level taxonomy built upon the GTDB taxonomy. The taxonomy of the species representatives sourced from the other 13 resources is determined using a custom approach based on the GTDB workflow (Chaumeil *et al.* 2022; Aroney *et al.* 2024) automatically assigning new taxon labels according to their Relative Evolutionary Divergence (RED) value. A Sylph (Shaw and Yu 2024) database and SingleM metapackage (Woodcroft *et al.* 2024) of GlobDB species representative genomes is available for taxonomic profiling of metagenomes. Additional genome metadata includes: completeness, contamination, and basic genome statistics calculated using CheckM2 (Chklovski *et al.* 2023); and whether the genome accession is linked to an isolate deposited in a major culture collection assessed via BacDive (Schober *et al.* 2025).

Furthermore, anvi'o (Eren *et al.* 2021) "contigs databases" of all GlobDB genomes are made available. These SQL databases contain the genome sequences and contextual data, including: gene calls for rRNA, tRNA, and protein coding genes as well as functional annotations of protein coding genes generated using Pfam (Mistry *et al.* 2021), COG (Galperin *et al.* 2021), KEGG (Kanehisa *et al.* 2023; Kananen *et al.* 2025), and dbCAN2 CAZymes (Zhang *et al.* 2018). GFF files with gene calls and COG IDs, fastA files with amino acid sequences of protein coding genes, and tab delimited files with functional annotations of each GlobDB genome are also available for download separately.

Finally, to be able to generate protein language model (pLM) embeddings of the GlobDB proteins, we clustered the 838,615,274 GlobDB amino acid sequences using a 40% identity threshold over 80% of the length of both sequences (Steinegger and Söding 2018), yielding 82,973,016 representatives of clusters larger than one sequence. We generated ProtT5-XL-U50 (Elnaggar *et al.* 2022) pLM embeddings for 82,972,511 of these cluster representatives, excluding proteins >9000 amino acids for technical reasons. Sequences, annotations, cluster membership, and pLM embeddings of this clustered sequence set are available for download.

Full methods for the GlobDB can be found on https://globdb.org/methods.

**Data availability**

The GlobDB is hosted on the Life Science Compute Cluster (LiSC) of the University of Vienna. The previous (220) and current (226) releases are available at https://globdb.org/. Currently available releases will be hosted for at least 10 years after publication. In addition, we intend to keep following the annual update schedule of the GTDB, and will keep previous releases available for reproducibility of work that used the GlobDB.

**Acknowledgements**

We thank Petra Pjevac for helpful discussion and critical review of the manuscript text. We thank Phil Hugenholz, Donovan Parks, and Pierre Chaumeil for early access to the GTDB species representatives, and Taylor Priest for sharing the mOTU species representatives. Computational results of this work have been achieved using the Life Science Compute Cluster (LiSC) of the University of Vienna. D.R.S., B.L.C., T.R., and M.W. were supported by the Cluster of Excellence grant "Microbiomes drive Planetary Health" of the Austrian Science Fund FWF (10.55776/COE7). N.P. and T.R. were supported by the project "DataLife – Data Infrastruktur for Life Sciences" of the "Digital Research Infrastructures" program of the Austrian Federal Ministry for Women, Science and Research. S.T.N.A. was supported by the EMERGE National Science Foundation (NSF) Biology Integration Institute (#2022070). B.J.W. was supported by Australian Research Council Future Fellowship (#FT210100521).

**Competing interests**

The authors declare that they have no competing interests